\documentclass{article}
\usepackage{iclr2020_conference,times}

\usepackage{amsmath,amsfonts,bm}

\def\eqref#1{equation~\ref{#1}}

\def\1{\bm{1}}

\DeclareMathAlphabet{\mathsfit}{\encodingdefault}{\sfdefault}{m}{sl}
\SetMathAlphabet{\mathsfit}{bold}{\encodingdefault}{\sfdefault}{bx}{n}

\usepackage{hyperref}
\usepackage{url}
\usepackage{graphicx}
\usepackage{gensymb}

\title{Assessing Robustness to Noise: \newline Low-Cost Head CT Triage}

\newcommand*\samethanks[1][\value{footnote}]{\footnotemark[#1]}

\author{Sarah M. Hooper\textsuperscript{1,4}\thanks{Equal contribution. For correspondence, please contact \texttt{smhooper@stanford.edu}}\ \ , Jared A. Dunnmon\textsuperscript{2}\samethanks\ \ , Matthew P. Lungren\textsuperscript{3}, \\ \textbf{Sanjiv Sam Gambhir\textsuperscript{3,4}, Christopher R\'e\textsuperscript{2}, Adam S. Wang\textsuperscript{3}\thanks{Equal contribution}\ \ \& Bhavik N. Patel\textsuperscript{3}\samethanks} \\
\textsuperscript{1}Department of Electrical Engineering, \textsuperscript{2}Department of Computer Science, \\ \textsuperscript{3}Department of Radiology, \textsuperscript{4}Molecular Imaging Program at Stanford\\
Stanford University\\
Stanford, CA 94305, USA 
}

\iclrfinalcopy
\begin{document}

\maketitle

\begin{abstract}
Automated medical image classification with convolutional neural networks (CNNs) has great potential to impact healthcare, particularly in resource-constrained healthcare systems where fewer trained radiologists are available. However, little is known about how well a trained CNN can perform on images with the increased noise levels, different acquisition protocols, or additional artifacts that may arise when using low-cost scanners, which can be underrepresented in datasets collected from well-funded hospitals. In this work, we investigate how a model trained to triage head computed tomography (CT) scans performs on images acquired with reduced x-ray tube current, fewer projections per gantry rotation, and limited angle scans. These changes can reduce the cost of the scanner and demands on electrical power but come at the expense of increased image noise and artifacts. We first develop a model to triage head CTs and report an area under the receiver operating characteristic curve (AUROC) of 0.77. We then show that the trained model is robust to reduced tube current and fewer projections, with the AUROC dropping only 0.65\% for images acquired with a 16x reduction in tube current and 0.22\% for images acquired with 8x fewer projections. Finally, for significantly degraded images acquired by a limited angle scan, we show that a model trained specifically to classify such images can overcome the technological limitations to reconstruction and maintain an AUROC within 0.09\% of the original model. 
\end{abstract}

\section{Introduction}

Computed tomography (CT) is a clinically important medical imaging modality with wide diagnostic scope. CT use has increased rapidly since its inception, with the percent of US emergency department visits involving CT rising from 2.8\% in 1995 to 13.9\% in 2007 \citep{larson2011national}. Globally, the demand for imaging systems such as CT is also likely to rise due to the increasing global burden of noncommunicable diseases and the increase in world population with growing numbers of older citizens \citep{ rohaya2011medical}. Automated CT triage could increase operational efficiency by ensuring patients with abnormal images are seen quickly, which holds promise particularly for healthcare systems with fewer trained radiologists. 

\begin{figure}[t]
\begin{center}
\includegraphics[width=.9\textwidth]{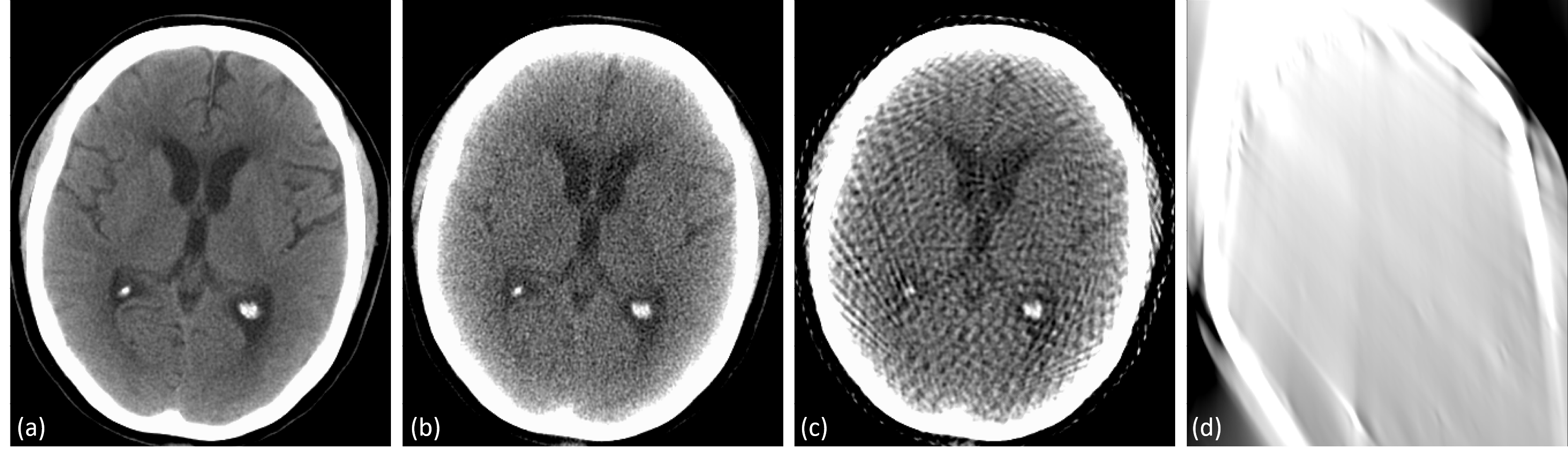}
\end{center}
\caption{Changing image acquisition results in changes to the noise characteristics of an image. (a) Head CT slice from the original dataset; (b) simulated 8x reduction in x-ray tube current; (c) simulated 8x reduction in number of projections; (d) simulated 1/8 angle scan.}
\end{figure}

Recently, convolutional neural networks (CNNs) have shown great potential for classifying medical images \citep{litjens2017survey}. As imaging systems and CNNs are deployed to more regions of the world, it is critical to understand how models perform with different imaging technology. However, much past work developing CNNs has been done with images collected from well-funded hospitals. This focus results in unknowns about how CNNs transfer to data collected from low-cost scanners, which may be more prevalent in resource-constrained healthcare systems. These unknowns are concerning as low-cost scanners can be more susceptible to image noise and artifacts unfamiliar to a trained CNN. For instance, low-cost CT may use low x-ray tube current, which increases image noise (Fig. 1(b)) but is less expensive, places lower demands on electrical power, and results in a longer x-ray tube lifespan than high tube current \citep{kalender1999dose}. Even though high tube current results in lower noise, these factors can be prohibitive in countries with limitations on electrical power, healthcare funding, and access to supply chains. CT scanners can also be made more affordable by simplifying scanner design. Some proposed low-cost CT scanners use reduced number of projections and limited angle scans to increase affordability \citep{liu2014top}. However, reduced projection and limited-angle acquisition result in distinctive image artifacts (Fig. 1(c,d)). 

Thus, to deploy CNNs with low-cost scanners, it is important to evaluate --- and retrain, if needed --- models with representative data. In this work, we evaluate the effect of (1) reducing x-ray tube current, (2) lowering the number of projections, and (3) acquiring a limited angle scan on the performance of a CNN trained to triage head CTs. As described above, these changes are likely to arise with low-cost CT scanners. However, representative data is limited, making it difficult to study how robust trained CNNs are to these changes or to train new CNNs to classify images from low-cost machines. In response, we partner with industry to use state-of-the-art simulation tools to generate large, synthetic datasets that allow us to systematically explore model performance with changes to acquisition protocols. Specifically, we collect a dataset of 9,776 head CTs from our institution's PACS and produce realistic simulations of those images as if they had been acquired with 4x, 8x, and 16x reduced tube current, number of projections, and acquisition angle. Although some of these changes --- particularly the limited angle scans and 16x reductions --- are severe, we were interested in exploring a wide range of degredations, including those which may enable new CT architectures (e.g. fixed source designs).

We begin by training a CNN to triage head CTs and report a 0.77 area under the receiver operating characteristic curve (AUROC). Then, we evaluate the triage model on the nine simulated datasets. We find the model is remarkably robust to some noise sources, maintaining an AUROC within 0.65\% of the original data for images acquired with 16x reduced tube current and 0.22\% for 8x fewer projections, but fails to accurately classify limited angle scans. We show that for applications like limited angle acquisition that lie outside the operating point of existing models, CNNs trained on the noisy data may overcome such technical limitations by training a model that classifies 1/4 angle scans within 0.09\% AUROC points of the original model on the unaltered images. This work elucidates how trained CNNs behave with suboptimal data, uncovers directions for automated triage with lower-cost imaging technology, and demonstrates the potential of simulated data as a means of evaluating CNNs along many image acquisition parameters that may not be otherwise accessible.

\section{Methods}

9,776 non-contrast head CTs were used in this study; dataset details can be found in the Appendix. 

\begin{figure}[t]
\begin{center}
\includegraphics[width=\textwidth]{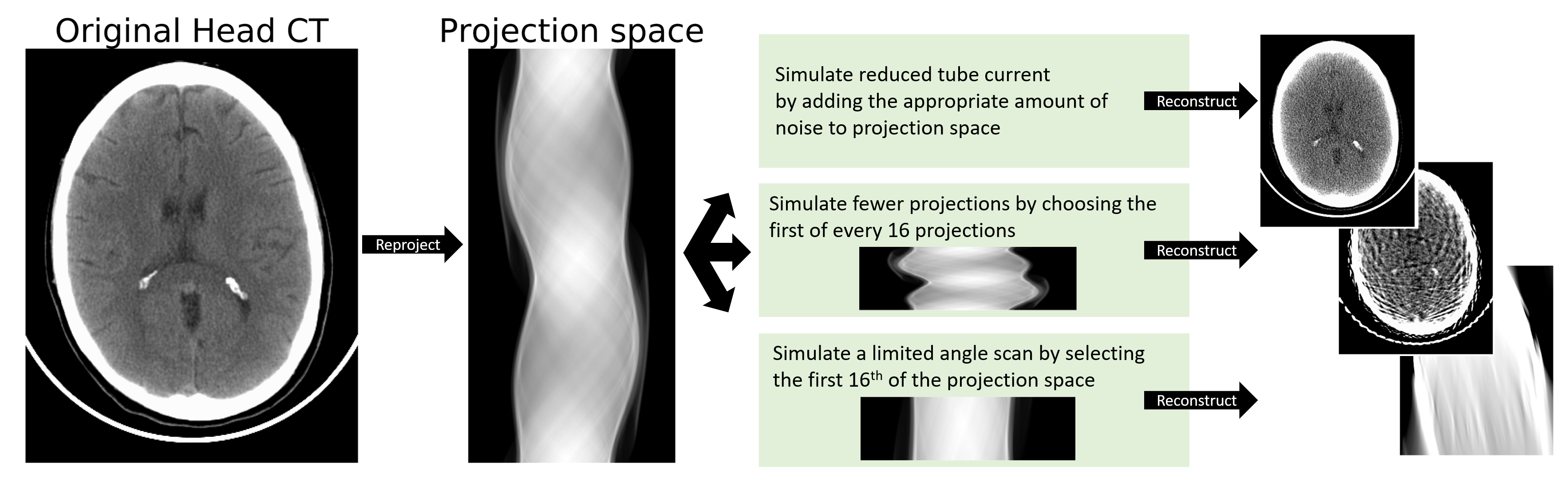}
\end{center}
\caption{Noise simulation process for 16x tube current, projection, and angle reduction.}
\end{figure}

\subsection{Model training procedure}

Using the Adam optimizer, we trained a 121-layer DenseNet, which was modified to take the single-channel 3D image volume as input and produces a normal/abnormal label as output \citep{huang2017densely}. Weights were initialized using the He initialization \citep{he2015delving}. Random horizontal flips, zooms, and rotations were applied to augment the training data. Coarse hyperparameter search was used to select the optimizer's hyperparameters, resulting in a learning rate of 1e-4 set to decay by 0.1 when the validation loss plateaued for 10 epochs and a weight decay of 1e-6. All other parameters were left at defaults. Batch size was one due to memory limitations; we accumulated gradients for 16 steps before backpropagation. Each model was trained for 50 epochs. All experiments were performed in Pytorch v1.2 on either a NVIDIA Tesla P100 or TITAN RTX GPU.

\subsection{Synthetic Dataset Generation}

To study model performance as tube current, projections, and acquisition angle changed, we created synthetic datasets by injecting noise into all images in our dataset.  We studied a 4x, 8x, and 16x reduction for each parameter. By using synthetic datasets, there was no need for additional image labeling, and it was possible to systematically explore many image parameters over large test sets. Past work has shown that simulating noise in projection space instead of in reconstructed image space is more accurate, where projection space represents the original sensor data collected by the imaging system prior to reconstruction (Fig. 2) \citep{benson2010synthetic}. Thus, we reprojected the CT scans back into projection space using GE’s CatSim software, which has been validated as a high fidelity simulation system \citep{de2007catsim}; still, validating our study on real world data is a promising direction for future work. First, the original CTs were used to create voxelized phantoms made of water and bone in which densities scaled with CT number. Then, we used CatSim to simulate axial scans similar to the original acquisition, including the system geometry, 120 kV spectrum, bowtie filter, and realistic beam hardening. All original scans were simulated to have 984 projections with the same angular range and start angle. To generate the 4x reduced projection scans, the first of every 4 projections were taken in projection space. The 4x limited angle scans were generated by taking the first quarter of all projections. The 4x reduced tube current was simulated by adding Gaussian white noise to projection space with variance accounting for the reduced transmitted photon flux \citep{benson2010synthetic}; original scans were acquired with X-ray tube current in the range of 123-440 mA. Note that in the reduced tube current scans, only the quantum noise was modeled, leaving the nonlinear effect of electronic noise as subject for future work. Similar procedures were followed to generate the 8x and 16x datasets (Fig. 2). Finally, the modified sinograms were reconstructed using CatSim.

\section{Results and Discussion}

\subsection{Head CT Triage Model}

\begin{figure}[t]
\begin{center}
\includegraphics[width=.7\textwidth]{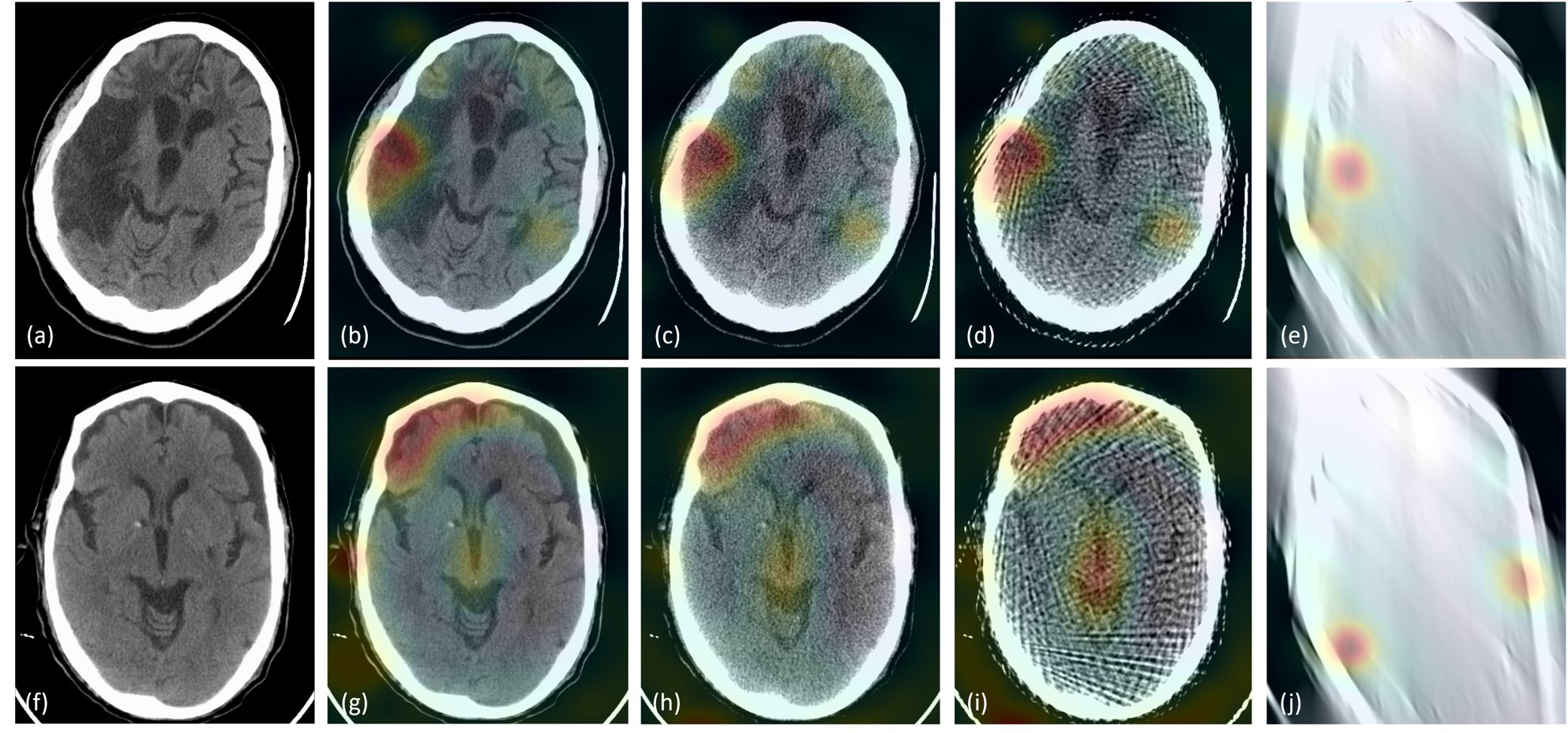}
\end{center}
\caption{(a)-(e) and (f)-(j) depict axial slices from two patients; (a), (b), (f) and (g) show the original image, (c) and (h) are the 8x reduced tube current images, (d) and (i) are the 8x reduced projection images, and (e) and (j) are the 1/8 angle scans. Overlaid on each image (b)-(e) and (g)-(j) is the class activation map of the abnormal class computed by the original model on the image.}
\end{figure}

Our first aim was to train a head CT triage model over the original dataset to establish baseline triage performance on images without any added noise. We evaluate our final model by computing the AUROC for three models trained as described in Section 2.1 with different random seeds. The final model achieves a mean AUROC value of 0.771 on the test set (std = 0.006). The class activation maps (CAMs) for the abnormal class on two example axial slices are shown in Fig. 3(b,g), exhibiting clear activation at sites of abnormalities. 

\subsection{Performance of Original Model on Noisy Datasets}

Our next aim was to evaluate the impact of low tube current, fewer projections, and limited angle acquisition on the trained triage model. Using the model trained on the original data, we classified the test images in the nine synthetically altered datasets and computed the AUROCs (left three panels of Fig. 4). The performance of the model on all levels of tube current reduction remains remarkably stable, achieving a 0.89 Cohen’s kappa coefficient on all tube current reduction datasets compared to the original dataset. Despite the increase in image noise resulting from reducing tube current by 16x, mean AUROC drops by only 0.65\%. This result is promising for model transfer between systems that operate with different tube current, and offers an exciting prospect for reducing patient radiation dose. Note that while the noise resulting from reduced tube current could be diminished by increasing scan time, longer scans increase motion artifacts and reduce scanner throughput. As mentioned, one limitation here is that the nonlinear effect of electronic noise is not modeled; however, the similarity of the performance of the model for tube current reduction (which will be impacted by modeling the electronic noise) and projection reduction (which will not be impacted by modeling the electronic noise) is promising for these results holding even once electronic noise is added to the model. Additionally, analysis of model performance over different subsets of the data would better elucidate the limitations of models over reduced tube current and reduced projection data. 

Limited angle scans severely impact the performance of the trained model, which achieves less than 0.03 Cohen's kappa coefficient on all limited angle datasets compared to the model's performance on the original dataset. The mean sensitivity on the 1/4 angle scans was 0.98 while the mean specificity was 0.04, revealing that nearly all limited angle images were classified as abnormal by the original model. In Fig. 3, we show the same axial slice of two patients' CT scans with the CAMs computed by the original model on the noisy images. The CAM takes high values in near-identical locations on the original, 8x reduced tube current, and 8x fewer projection images, providing intuitive confirmation that the model is finding the same abnormalities despite the additional artifacts. However, the CAM fails to locate those abnormalities on the limited angle scans. 

\begin{figure}[t]
\begin{center}
\includegraphics[width=\textwidth]{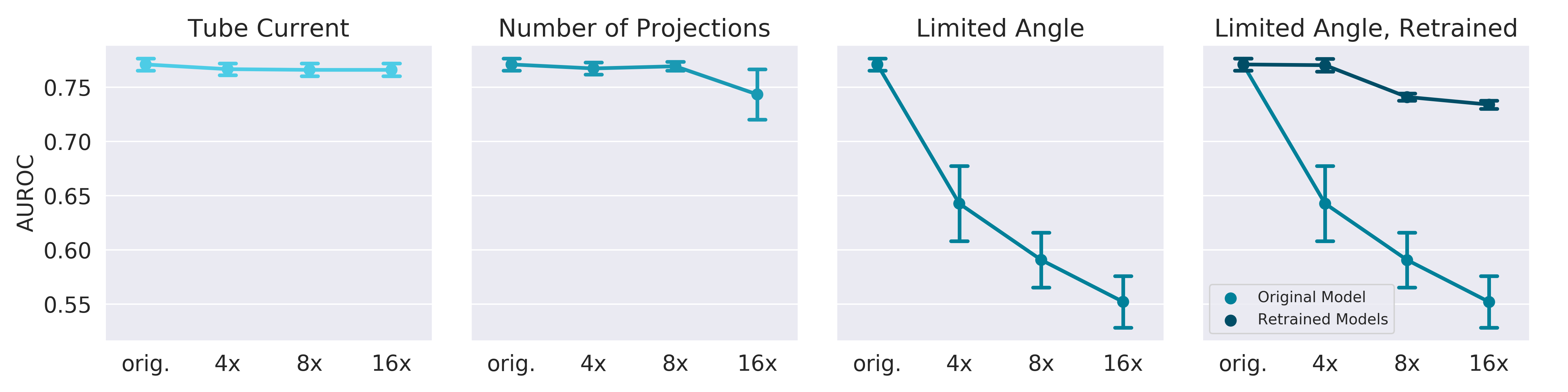}
\end{center}
\caption{Mean AUROC for reduced current, reduced projection, and limited angle scans. Y axes are shared; error bars show standard deviation of three models trained with different random seeds.}
\end{figure}

\subsection{Performance of Models Trained on Limited Angle Scans}

Given the low performance of the original model on the limited angle scans, our final aim was to investigate the performance of a model trained specifically on these images. For each limited angle dataset, a new model was trained following the procedure in Section 2.1. The performance of the limited angle models compared to the original model on the limited angle datasets is shown in the rightmost panel of Fig. 4. Using models trained specifically for triage of limited angle scans, we substantially improve performance and recover nearly the same performance for the 1/4 angle scans as the original model on the full angle scans. Although limited angle scans are near uninterpretable to the human eye, the new models are able to maintain high performance. While this high performance is at first surprising, it makes sense in the context of past work in sinogram inpainting and the simplicity of the inverse radon transform; much of the information needed for triage is still available in the limited angle sinogram, even though it is insufficient to support reconstruction via filtered back projection. Since the cost and technical requirements of a limited angle CT scanner may be significantly lower than standard CT, these results are promising for the ability of machine learning models to overcome technological limitations. Still, careful consideration should be given to the impact on human readers, perhaps pairing a detection network with a sinogram inpainting method.

\section{Conclusion}

Though automated medical image classification holds promise in resource-constrained healthcare, little work has investigated how models perform with suboptimal data. This work evaluates the impact of low tube current, fewer projections, and limited angles on trained CNNs using realistic simulations. We show that a model trained to triage head CTs remains stable despite decreased tube current and number of projections and that CNNs can triage limited angle scans, suggesting that machine learning models may help overcome significant hardware limitations to image analysis.

In this work we use simulations to study noise from low-cost scanners, which enables systematic evaluation over large datasets without increasing labeling demand. However, studying variations in acquisition protocol using synthetic data is relevant when considering model deployment in any healthcare system. Different institutions often have differing acquisition protocols, with noise levels adjusted to suit the needs of their healthcare practitioners. However, robustness tests over acquisition protocol and noise level are rarely reported. Thus, the line of work presented in this study is relevant for model testing prior to deployment within any healthcare system. Finally, learning directly in sinogram space instead of reconstructed image space is an interesting future study that may also be pursued with synthetic data.

\subsubsection*{Acknowledgements}
This study was in part funded by GE Healthcare. Sarah Hooper is supported by the Fannie and John Hertz Foundation, the National Science Foundation Graduate Research Fellowship under Grant No. DGE-1656518, and as a Texas Instruments Fellow under the Stanford Graduate Fellowship in Science and Engineering. We also gratefully acknowledge the support of DARPA under Nos. FA86501827865 (SDH) and FA86501827882 (ASED); NIH under No. U54EB020405 (Mobilize), NSF under Nos. CCF1763315 (Beyond Sparsity), CCF1563078 (Volume to Velocity), and 1937301 (RTML); ONR under No. N000141712266 (Unifying Weak Supervision); the Moore Foundation, NXP, Xilinx, LETI-CEA, Intel, IBM, Microsoft, NEC, Toshiba, TSMC, ARM, Hitachi, BASF, Accenture, Ericsson, Qualcomm, Analog Devices, the Okawa Foundation, American Family Insurance, Google Cloud, Swiss Re, the HAI-AWS Cloud Credits for Research program, and members of the Stanford DAWN project: Teradata, Facebook, Google, Ant Financial, NEC, VMWare, and Infosys. The U.S. Government is authorized to reproduce and distribute reprints for Governmental purposes notwithstanding any copyright notation thereon. Any opinions, findings, and conclusions or recommendations expressed in this material are those of the authors and do not necessarily reflect the views, policies, or endorsements, either expressed or implied, of DARPA, NIH, ONR, or the U.S. Government.

\clearpage
\bibliography{iclr2020_conference}
\bibliographystyle{iclr2020_conference}

\section*{Appendix}

\subsection*{Dataset}

We began by collecting 10,000 non-contrast head CTs acquired between 2001 and 2014 from our institution’s PACS under an approved IRB. There were between 28 and 56 axial slices per scan with 5 mm axial reconstructed slices of 512x512 voxels each. Each volume was padded with zeros to 56 axial slices. The mean and standard deviation over all training images were used to normalize each image. We randomly split the data into 80\% training, 10\% validation, and 10\% test sets. After filtering to ensure no distinct scans from the same patient were in the train and validation or test sets, 7,856, 936, and 984 CTs were left in the train, validation, and test sets respectively, totalling 9,776 total images in our dataset. Each exam was prospectively labeled by a radiologist at the time of interpretation as normal or abnormal, using additional data available to the clinician  (e.g. patient history, age, symptoms). Although not labeled for inter-rater reliability, the labels that we used to train the model were the same ones acted upon by the healthcare team at the time of patient care. The dataset was not filtered by pathology, so it included a diverse set of abnormalities such as hemorrhage, stroke, fracture, fluid collection, and extracranial injuries. Over all images in the dataset, 55.22\% were abnormal.

\subsection*{Related Work}

Much past work using deep learning to address noise in medical images has focused on denoising images. For example, \citet{wurfl2018deep} use deep residual learning to reconstruct CTs from 180\degree\ limited angle projection data, and \citet{chen2017low} use neural networks to reduce noise in low tube current CTs. However, denoising is fundamentally different from our work described here. In denoising networks, the training and test sets consist of images with noise from a known source. In this work, we consider the consequences of a deployed model processing images with different noise characteristics than those of the training set. Some previous work has explored the robustness of CNNs to adversarial examples for medical image classification \citep{paschali2018generalizability}. However, the adversarial examples do not reflect realistic medical image noise. In the area of head CT triage, much past work has used CNNs to identify a particular pathology \citep{arbabshirani2018advanced}. There have been some studies using CNNs to classify a range of cranial abnormalities \citep{chilamkurthy2018deep}, but these studies are limited by available multiclass training labels and cannot be used to triage pathologies not specifically labeled in their training set. Finally, \citet{titano2018automated} present a general head CT anomaly detection CNN. However, this network was weakly supervised using noisy labels generated from a natural language processing (NLP) model. Though an efficient means of labeling, their results show that the CNN achieves 0.88 AUROC against the NLP-generated labels but drops to 0.73 AUROC against the radiologist-given labels. 

\end{document}